\documentclass{article}
\usepackage{graphicx,psfig,epsfig}
\def\title{\begin{center}\Large\bf}
\def\author(s){\vspace{0.3cm}\large\rm}
\DeclareRobustCommand{\ion}[2]{%
\relax\ifmmode
\ifx\testbx\f@series
{\mathbf{#1\,\mathsc{#2}}}\else
{\mathrm{#1\,\mathsc{#2}}}\fi
\else\textup{#1\,{\mdseries\textsc{#2}}}%
\fi}
\newcommand{\ha}{\rm H$\alpha$}

\newcommand{\HII}{\ion{H}{ii}}
\newcommand{\hnii}{{\rm H}$\alpha+[$\ion{N}{ii}$]$}

\newcommand{\sii}{$[$\ion{S}{ii}$]$}

\newcommand{\oiii}{$[$\ion{O}{iii}$]$}
\newcommand{\snr}{G 69.4+2.4}
\newcommand{\vel}{km s$^{-1}$}
\newcommand{\et}{{et al.}}
\newcommand{\dens}{cm$^{-3}$}
\def\arcmin{\hbox{$^\prime$}}
\def\arcsec{\hbox{$^{\prime\prime}$}}
\def\degr{\hbox{$^\circ$}}
\def\text{\end{center}}

\begin{document}

\title
Optical observations of the supernova remnant AX~J2001+3235

\author(s)
 F. Mavromatakis$^{\rm 1}$, P. Boumis$^{\rm 1}$, E.V. Paleologou$^{\rm 2}$

$^{\rm 1}${\it University of Crete, P.O Box 2208, GR-71003, Heraklion, Greece}\\
$^{\rm 2}${\it Foundation for Research and Technology-Hellas, P.O. Box 1527,\\
GR-71110 Heraklion, Greece}\\
\text

\vspace{0.3cm}

\large

\section*{Abstract}
We present results from a first analysis of optical data on \snr\  
obtained with the 0.3 m and 1.3 m telescopes at Skinakas Observatory. 
Filamentary emission is detected in the \oiii 5007 \AA\  and \hnii\ 
images oriented in the south--west, north--east direction.
The \hnii\ images reveal a complex network of filaments, 
while the \oiii\ images show a single filament. 
Low resolution spectra were also obtained which identify the observed emission
as emission from shock heated gas.

\section{Introduction}
Asaoka \et\ (1996) reported the detection of extended soft X--ray emission 
in the ROSAT All--Sky survey. This X--ray emission is located in Cygnus 
and has an angular extent of $\sim$3\degr $\times$ 1\degr. The authors did not 
find any strong source in the area which could power this X--ray arc.  
Subsequent pointed observations by ROSAT and ASCA (Yoshita et al. 2000) 
showed that the emission was thermal with a temperature of 0.4 keV. 
The hydrogen column density was found in the range of 1--3 $\times$ 
10$^{21}$ cm$^{- 2}$, similar to the column density towards the nearby 
remnant CTB~80 (Mavromatakis \et\ 2001). However, the spectral analysis 
was based only in the core of the X--ray arc which has an extent 
of $\sim$1\degr. Assuming an explosion energy of 10$^{51}$ erg and the 
Sedov--Taylor solution, Yoshita \et\ (2000) proposed a radius of 22 pc, 
an age of 37000 yrs and a hydrogen density of 0.06 \dens. 
Recently, Lu \& Aschenbach (2001) proposed that the whole 3\degr\ X--ray arc 
is the supernova remnant G~69.4+1.2 and not only the 1\degr\ core since the 
ROSAT survey data did not show any spectral variations across the whole 
extent of the arc.  
\section{Observations}
Deep CCD imaging observations were performed with a 1024 $\times$ 1024 Site CCD 
which in conjuction with the 0.3 m telescope at Skinakas Observatory 
resulted in a pixel scale of 4 arcseconds. Higher resolution images 
were obtained with the 1.3 m telescope at a pixel scale of 0.5 arcsec. The
filters used during these imaging observations were isolating the emission 
lines of \oiii, \sii\ and \hnii. 
The 0.3 m telescope observations are characterized by exposure times around 
7000 sec, while the 1.3 m telescope imaging observations 
are characterized by exposure times of 1800 sec. 
The astrometric solution calculated for every image utilized the Hubble
Guide star catalogue.  
Deep low resolution spectra were also taken with the 1.3 m telescope 
with a total exposure time of 7800 sec. The spectra cover the range 
of 4750 \AA\ -- 6816 \AA\ at a scale of $\sim$ 1 \AA\ per pixel.
\begin{figure}
\centering
\mbox{\epsfclipon\epsfxsize=2.8in\epsfbox[72 166 540 626]{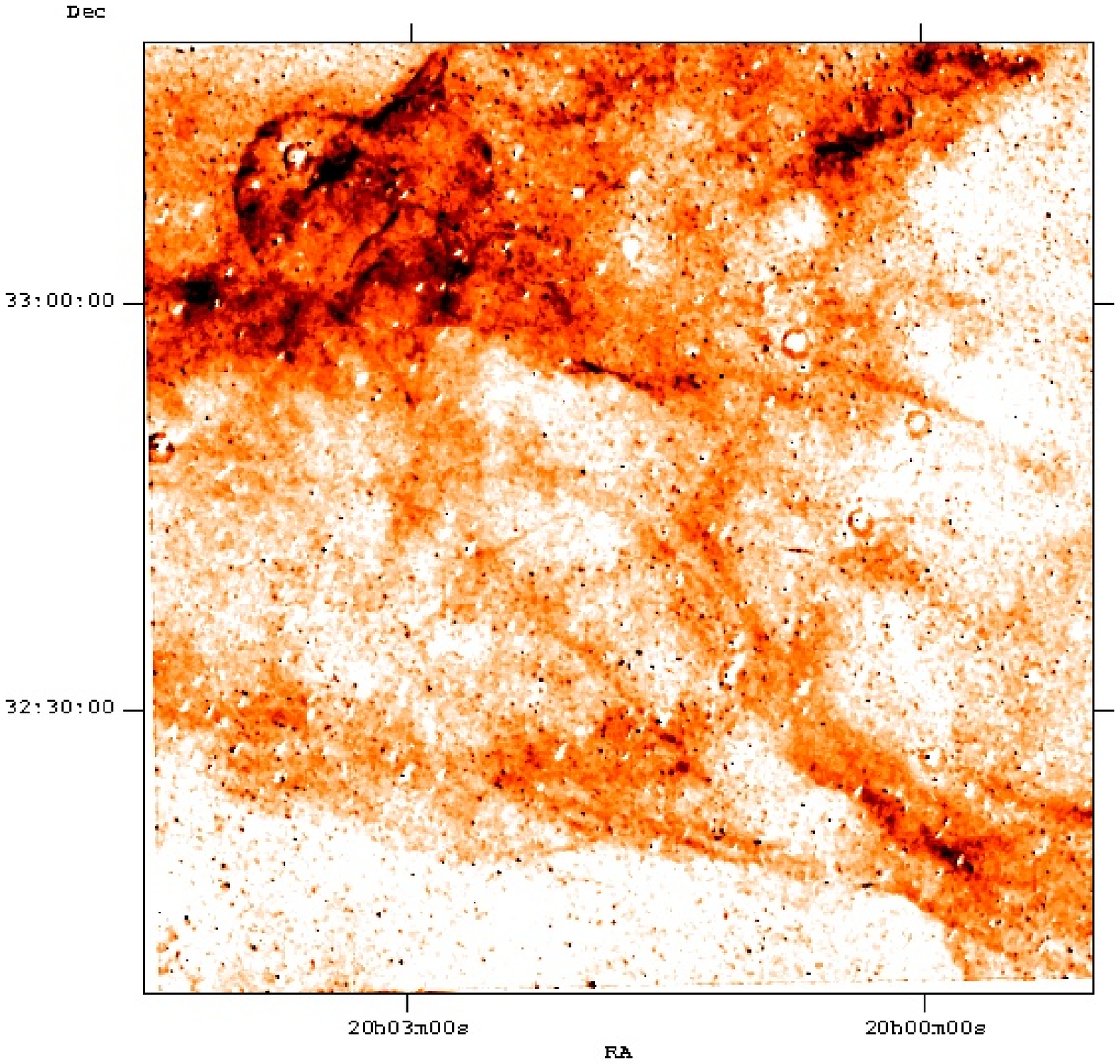}}
\\ {{\bf Figure 1.} The area round \snr\ in the low ionization lines of \sii. 
A strong \HII\ region (LBN 159; Lynds 1965) is present in the north, while 
filamentary and diffuse emission is detected in the south, south--west.}
\label{fig1}
\end{figure}
\section{Current preliminary results}
The \hnii\ and \sii\ (Fig. 1) images reveal the presence of both filamentary 
and diffuse structures extending further to the west, south--west.
Emission is also detected in the area of G~69.7+1.0 but it is not clear 
if it belongs to it or to  G~69.4+1.2. The extended emission to the north 
edge of our field belongs to LBN 159 (Lynds 1965). 
The \sii\ emission is quite strong relative to \ha\ since we estimate a 
\sii/\ha\ ratio around 0.9, while this ratio drops to $\sim$0.4 within LBN 159. 
The \oiii\ image (Fig. 2) shows that LBN 159 also emits in this line, while 
only a $\sim$5\arcmin\ long filament is detected elsewhere in the field.
The fields in Figs. 1 and 2 are $\sim$ 70\arcmin$\times$ 70\arcmin, while the 
higher resolution images cover an extent $\sim$ 8.5\arcmin$\times$ 8.5\arcmin\ 
(Fig. 3).
The higher resolution images show that the projected thickness of the 
\oiii\ filament to be $\sim$ 5\arcsec, while the \hnii\ image reveals a 
more complex structure. 
The deep long slit spectra taken at the position of the \oiii\ filament 
show that we have observed emission from shock heated gas since we measure a
\sii\ to \ha\ ratio greater than 1.
The ratio of the two sulfur lines at 6716 \AA\ and 6731 \AA\ suggests very 
low electron densities, below $\sim$120 \dens. 
The strong \oiii\ emission relative to \ha\ implies either fast shocks
($\sim$120 \vel) or the presence of incomplete recombination zones. 
\begin{figure}
\centering
\mbox{\epsfclipon\epsfxsize=2.8in\epsfbox[72 181 540 611]{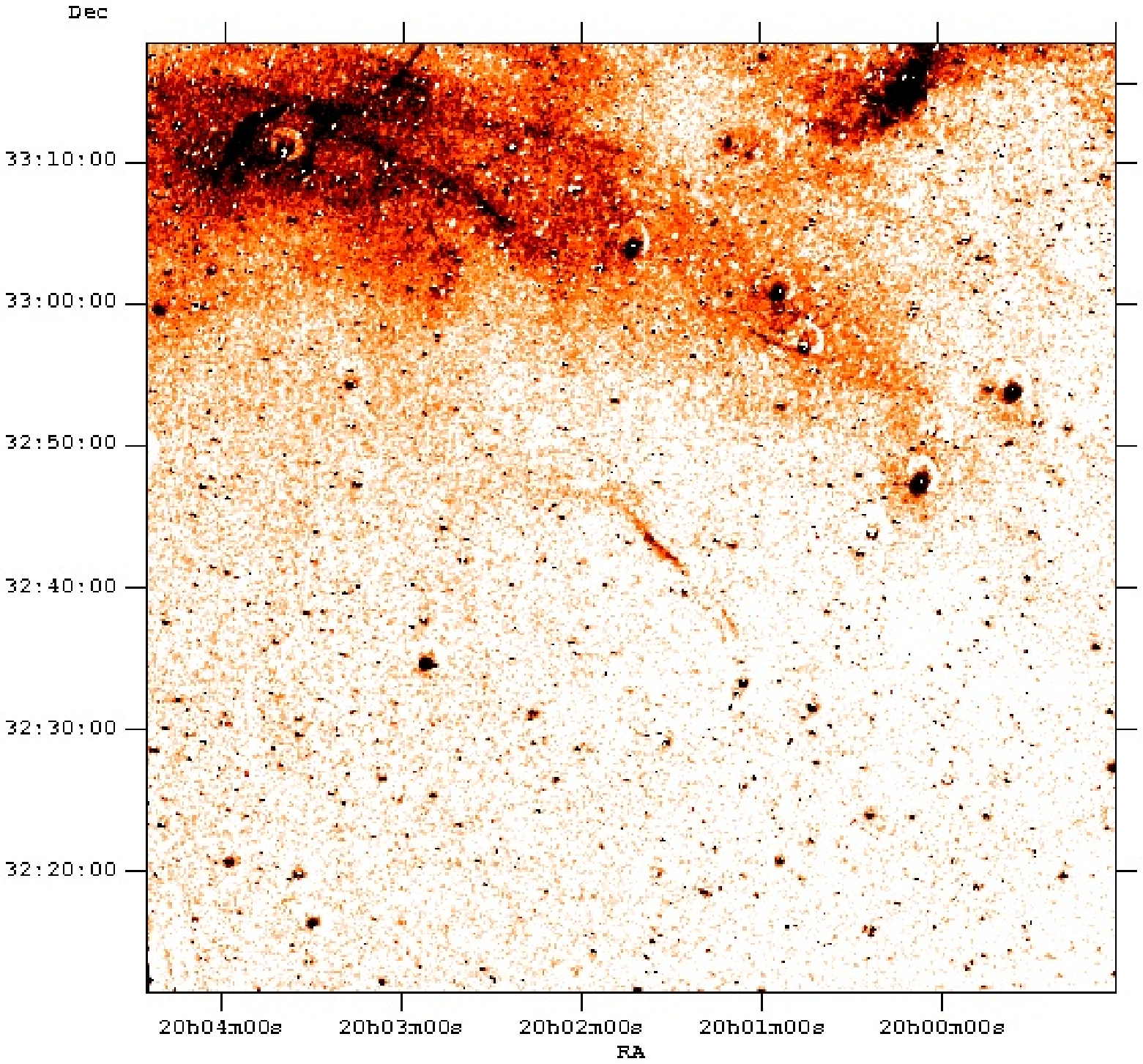}}
\\ {{\bf Figure 2.} The same field as in Fig. 1 is shown in the medium 
ionization line of \oiii 5007\AA. The \HII\ emission is significantly suppressed, 
while a filament is present near the center of our field.\ }
\label{fig2}
\end{figure}
The radio emission, although weak, is very well correlated with the optical 
data. Eventhough good quality radio data at other wavelengths are not 
available, we believe that the nature of the radio emission must be 
non--thermal given the strong sulfur line emission relative to \ha\ 
measured in the flux calibrated images.
The nature of the optical emission and possibly that of the radio, the 
common orientation of the observed emission (south--west to north--east) 
and the fact that the outer soft X--ray contours roughly trace the south 
boundaries of the optical and radio emission would favor the association 
of the optical radiation with the emission detected in the ROSAT All--Sky 
survey.
\begin{figure}
\centering
\mbox{\epsfclipon\epsfxsize=2.8in\epsfbox[72 278 540 514]{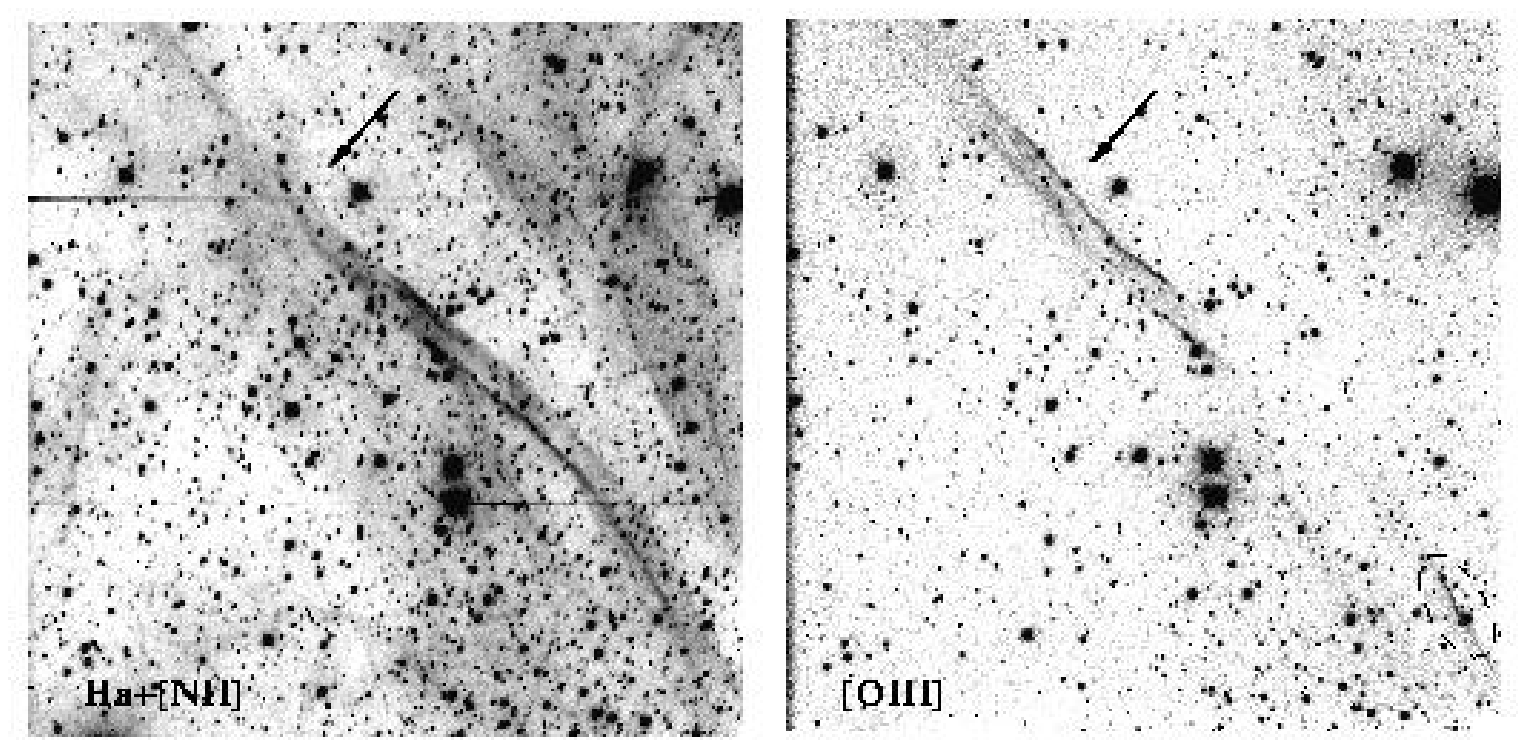}}
\\ {{\bf Figure 3.} The area around the prominent \oiii\ emission.  
The \hnii\ image is shown to the left, while the \oiii\ image of the same 
area is shown to the right. The arrow points to the slit center, while 
the dashed ellipse delinates the presence of filamentary \oiii\ emission 
in the south--west edge of our field.}
\label{fig3}
\end{figure}
\section*{Acknowledgments}
Skinakas Observatory is a collaborative project of the
University of Crete, the Foundation for Research and Technology-Hellas
and the Max-Planck-Institut fur Extraterrestrische Physik.

\section*{References}
Asaoka I, Egger, R. and Aschenbach B. 1996, MPE report 263
Lu F. J. and Aschenbach B. 2001, to be submitted \\
Mavromatakis et al. 2001, to be submitted \\
Yoshita K., Miyata E. and Tsunemi H. 2000, PASP 52, 867 \\

\end{document}